%% file: main.tex
\documentclass[10pt,a4paper,oneside,fleqn]{article}

\usepackage[utf8]{inputenc}

\usepackage{graphicx}
\graphicspath{{figure/}}
\usepackage[footnotesize,bf,hang]{subfigure}
\usepackage[small,bf]{caption}
\usepackage{amsmath,amsfonts,amssymb,amsthm}
\usepackage{latexsym}

\usepackage[a4paper,top=3.cm,bottom=3.5cm,left=3.0cm,right=3.0cm]{geometry}
\usepackage{setspace}

\usepackage{titling}
\usepackage{sectsty}
\sectionfont{\large}
\subsectionfont{\normalsize}

\usepackage{authblk}

\newcommand{\sub}[1]{$_{\text{#1}}$}
\newcommand{\up}[1]{$^{\text{#1}}$}

\makeatletter
\def\@maketitle{%
  \newpage
  \null
  \begin{flushleft}
  \let \footnote \thanks
    {\LARGE  \@title \par}%
    \vskip 1.5em%
    {\normalsize
      \@author
      \par}%
  \end{flushleft}
  \par
  \vskip 1.5em}
\makeatother

\title{\textbf
{Two-dimensional TiO\sub{x} nanostructures on Au(111): a Scanning Tunneling Microscopy and Spectroscopy investigation}}
\date{}

\author[a]{F. Tumino}
\author[a]{P. Carrozzo}
\author[a]{L. Mascaretti}
\author[a]{C. S. Casari}
\author[a]{M. Passoni}
\author[b]{S. Tosoni}
\author[a]{C. E. Bottani}
\author[a]{A. Li Bassi}
\affil[a]{\footnotesize{Department of Energy and NEMAS (Center for NanoEngineered MAterials and Surfaces), Politecnico di Milano, Via Ponzio 34/3, I-20133, Milan, Italy}}
\affil[b]{Department of Materials Science, Università di Milano Bicocca, via Roberto Cozzi 55, 20125 Milano, Italy}

\begin{document}

\maketitle

\begin{abstract}
\noindent We investigated the growth of titanium oxide two-dimensional (2D) nanostructures on Au(111), produced by Ti evaporation and post-deposition oxidation. Scanning tunneling microscopy and spectroscopy (STM and STS) and low-energy electron diffraction (LEED) measurements characterized the morphological, structural and electronic properties of the observed structures. Five distinct TiO\sub{x} phases were identified: the \emph{honeycomb} and \emph{pinwheel} phases appear as monolayer films wetting the gold surface, while nanocrystallites of the \emph{triangular}, \emph{row} and \emph{needle} phases grow mainly over the honeycomb or pinwheel layers. Density Functional Theory (DFT) investigation of the honeycomb structure supports a $(2\times 2)$ structural model based on a Ti-O bilayer having $\text{Ti}_2\text{O}_3$ stoichiometry. The pinwheel phase was observed to evolve, for increasing coverage, from single triangular crystallites to a well-ordered film forming a $(4\sqrt{7}\times 4\sqrt{7})R19.1^\circ$ superstructure, which can be interpreted within a moiré-like model. Structural characteristics of the other three phases were disclosed from the analysis of high-resolution STM measurements. STS measurements revealed a partial metallization of honeycomb and pinwheel and a semiconducting character of row and triangular phases.
\end{abstract}

\input{introduction}

\input{experimental}

\input{results}

\input{conclusions}

\input{references}
\end{document}

%% file: introduction.tex
\section{Introduction}
\indent Titanium dioxide is a model system in the surface science of oxide materials. The physical and chemical properties of rutile and anatase surfaces have been the focus of an ongoing research activity \cite{diebold}. Increasing interest has been recently devoted to the study of two-dimensional (2D) titania structures, encouraged by the promising potentialities attributed to 2D oxide systems for several advanced nanotechnology applications \cite{pacc}. The deposition of ultrathin TiO\sub{x} films and nanostructures has been performed on several metal substrates (e.g. Cu(001) \cite{finetti}, Mo(112) \cite{chen}, Pt(001) \cite{matsu}, Pt(111) \cite{sedona1}, Au(111) \cite{wu}), leading to the observation of novel few-layers, well-ordered titanium oxide phases which, due to confinement effects and the interaction with the substrate, exhibit structural, chemical and electronic properties different from those of bulk TiO\sub{2} surfaces. For instance, ultrathin TiO\sub{x} films on Pt(111) have been shown to adopt peculiar 2-D structures, such as the so-called \emph{kagomé}, \emph{wagon-wheel} (or \emph{pinwheel}) and \emph{zig-zag} phases \cite{sedona1,barcaro}. Such structures have also been used as templates for the ordered nucleation of Au nanoclusters \cite{sedona2}, being TiO\sub{2} supported gold particles a well investigated catalyst \cite{valden}. Also the ``inverse system'' TiO\sub{x}/Au(111) has been observed to show an effective catalytic activity in promoting the water-gas shift (WGS) reaction \cite{rodri}, thus encouraging experimental work on the synthesis and characterization of TiO\sub{x} structures supported by the Au(111) surface. After the oxidation of Ti clusters evaporated on Au(111), the formation of geometrically well-defined TiO\sub{2} crystallites with triangular or hexagonal shape was reported by Biener et al. \cite{biener} and Potapenko et al. \cite{pota}. Hexagonal TiO\sub{x} nanoparticles were also observed by Lauwaet et al. \cite{lauwaet2} who revealed a semiconducting n-type character. The evolution of TiO\sub{x} structures for increasing coverage was investigated by Wu et al. \cite{wu} who reported the observation of two different wetting films, namely the so-called \emph{honeycomb} and \emph{pinwheel} phases, in addition to triangular islands. Other works \cite{raga1,raga2,farstad}, based on chemical vapour deposition methods, allowed to observe different titania phases referred to as \emph{honeycomb}, \emph{pinwheel}, \emph{star} and TiO\sub{2}(B). The mentioned works reveal a remarkable sensitivity of the TiO\sub{x}/Au(111) system to the preparation procedure. For instance, in the case of PVD procedures based on Ti evaporation, several experimental factors (e.g. O\sub{2} partial pressure either during or after Ti deposition, post-deposition annealing temperature and duration), can be expected to play a key role in determining the morphology of the resulting structures.\\
\indent In this work we investigate in detail the growth of TiO\sub{x} nanostructures on Au(111) from sub-monolayer to 1.5 monolayers (ML). The preparation method is based on Ti atoms evaporation on the clean Au(111) surface followed by O\sub{2} exposure and UHV annealing. Morphological and structural nanoscale properties have been investigated by means of scanning tunneling microscopy (STM) and low-energy electron diffraction  (LEED) measurements. Our experiments allowed for the identification, at various coverages, of five different TiO\sub{x} phases, whose structural characteristics will be extensively discussed in sec. 3 in light of atomically resolved STM images. The honeycomb phase (sec. 3.1) has been the subject of a Density Functional Theory (DFT) investigation aimed to provide a better understanding of its structural and electronic properties, leading to the validation of a previously proposed structural model \cite{wu}. To gain insight into the electronic characteristics of the observed phases around the Fermi level we performed scanning tunneling spectroscopy (STS) measurements, whose results will be presented in sec. 3.4.

%% file: experimental.tex
\section{Experimental and theoretical methods}
The experiments were performed under ultra-high vacuum condition (base pressure $\sim 5\times 10^{-11}$ mbar) in a chamber equipped with a Variable-Temperature STM (Omicron Nanotechnology GmbH), LEED/AES instrumentation and standard tools for sample preparation. The (111) surface of a gold single crystal (MaTecK GmbH) was cleaned by 15 min of Ar\up{+} ion bombardment at 1 keV,  at a sample temperature of 850 K, which was kept for at least 30 min after the sputtering. This procedure was repeated until no contaminants were detected by means of AES and STM. The resulting $(22\times\sqrt{3})$ reconstruction on the Au(111) surface was confirmed by LEED and STM measurements prior to the TiO\sub{x} deposition. On the freshly prepared Au surface, we first deposited titanium at room temperature by electron-beam evaporation from a 99.99\% pure Ti rod (MaTecK GmbH). The amount of deposited Ti was estimated by analysing STM images of the sample prior to the oxidation process \cite{carrozzo}. In this paper, we express the surface coverage in equivalent monolayers (ML) of Ti, where 1 ML corresponds to the density of surface atoms of Au(111)-$(1\times 1)$, i.e. $1.4\times 10^{15}$ cm\up{-2}. The deposition rate was kept constant at 0.2 ML/min and Ti coverage was varied from 0.25 ML to 1.5 ML by varying the deposition time. To obtain titanium oxide structures, after Ti deposition, the sample at room temperature was exposed to $10^{-6}$ mbar O\sub{2} pressure for 500 s and then annealed at 850 K for either 15 min or 25 min whether Ti coverage was respectively below or above 1 ML, in order to achieve better ordered oxide structures. \\ 
\indent Constant-current STM images were taken at room temperature typically using sample-tip bias voltage in the range $-2$ V, $+2$ V (referred to the sample) and tunneling current in the range 0.2 - 2 nA. Differential conductivity curves were acquired at 100 K by means of a lock-in amplifier applying a modulation of 40 mV\sub{rms} at 8 kHz. For each of the observed titania structures several STS curves (at least 20, discarding those exhibiting high noise or electronic instability artifacts) were collected and then averaged to get a single spectrum containing global (i.e. spatially averaged) information on the electronic properties of the particular TiO\sub{x} structure under investigation. To properly extract such information, we applied a normalization method originally proposed by Ukraintsev \cite{ukr} and then further elaborated by Passoni \textit{et al.} \cite{passoni}, which essentially allows to normalize the differential conductivity curve over a properly optimized tunneling coefficient, in order to obtain data more closely related to the surface density of states (DOS). Moreover, STS measurements were repeated using tips of different material, namely W and Cr, in order to experimentally discriminate possible tip DOS features in the acquired spectra.\\
\indent All calculations have been performed with the simulation package VASP 5.2 \cite{kresse}. The core electrons are treated with the Projector Augmented Wave method \cite{blochl,kresse2}. O(2s, 2p), Ti(3s, 4s, 3p, 3d) and Au(6s, 5d) are considered as valence electrons and treated explicitly. The exchange and correlation energy is calculated according to the generalized gradient approximation (GGA) of the density functional theory, as formulated in the functional from Perdue, Burke and Ernzerhof (PBE) \cite{pbe}. We first relaxed the Au bulk lattice parameter by using a Gamma-centred mesh of $12\times 12\times 12$ K-points and a kinetic cutoff of 600 eV. The relaxed lattice constant (4.17 \AA) is in good agreement with the X-ray diffraction measurements (4.08 \AA) \cite{wyckoff}. The slab model of the (111) Au surface is cut from the relaxed bulk structure and constitutes of 5 layers. An empty space larger than 15 {\AA} is included in the cell to avoid any spurious interaction between replicas. The Ti\sub{2}O\sub{3} honeycomb structure is placed in a $2\times 2$ supercell of the Au(111) surface. The lattice parameters are kept fixed to the bulk relaxed values, i.e. the strain due to the lattice mismatch is transferred to the 2D supported phase. The structure relaxation is performed with a kinetic cutoff of 400 eV and a Gamma-centred net of $4\times 4\times 1$ k-points. The two bottom layers of the Au slab are kept fixed to the bulk lattice positions, while the three topmost Au layers and the Ti\sub{2}O\sub{3} atoms are completely relaxed. Truncation criteria of 10\up{-5} eV and 10\up{-2} eV/{\AA} are adopted for electronic and ionic relaxation, respectively. The simulation of STM images is performed within the Tersoff-Hamann approximation \cite{th}. An isosurface of 10\up{-3} {\AA\up{-3}} is displayed.

%% file: results.tex
\section{Results and discussion}
By investigating the evolution of the system for increasing coverage up to 1.5 ML, we were able to identify five different TiO\sub{x} phases. The structural characteristics of the \emph{honeycomb} phase are discussed in sec. 3.1, where, in addition to experimental data, the results of DFT calculations are reported. Sec. 3.2 focuses on the structure of the \emph{pinwheel} phase, which is interpreted in the light of a model based on the formation of a moiré coincidence pattern. In sec. 3.3 we discuss the structure of the other observed phases: \emph{triangular}, \emph{row} and \emph{needle}. In particular, we report atomically resolved images of row and needle crystallites, for which a detailed structural characterization is still lacking. 
\begin{figure}[htbp]
\centering%
\includegraphics[scale=0.6]{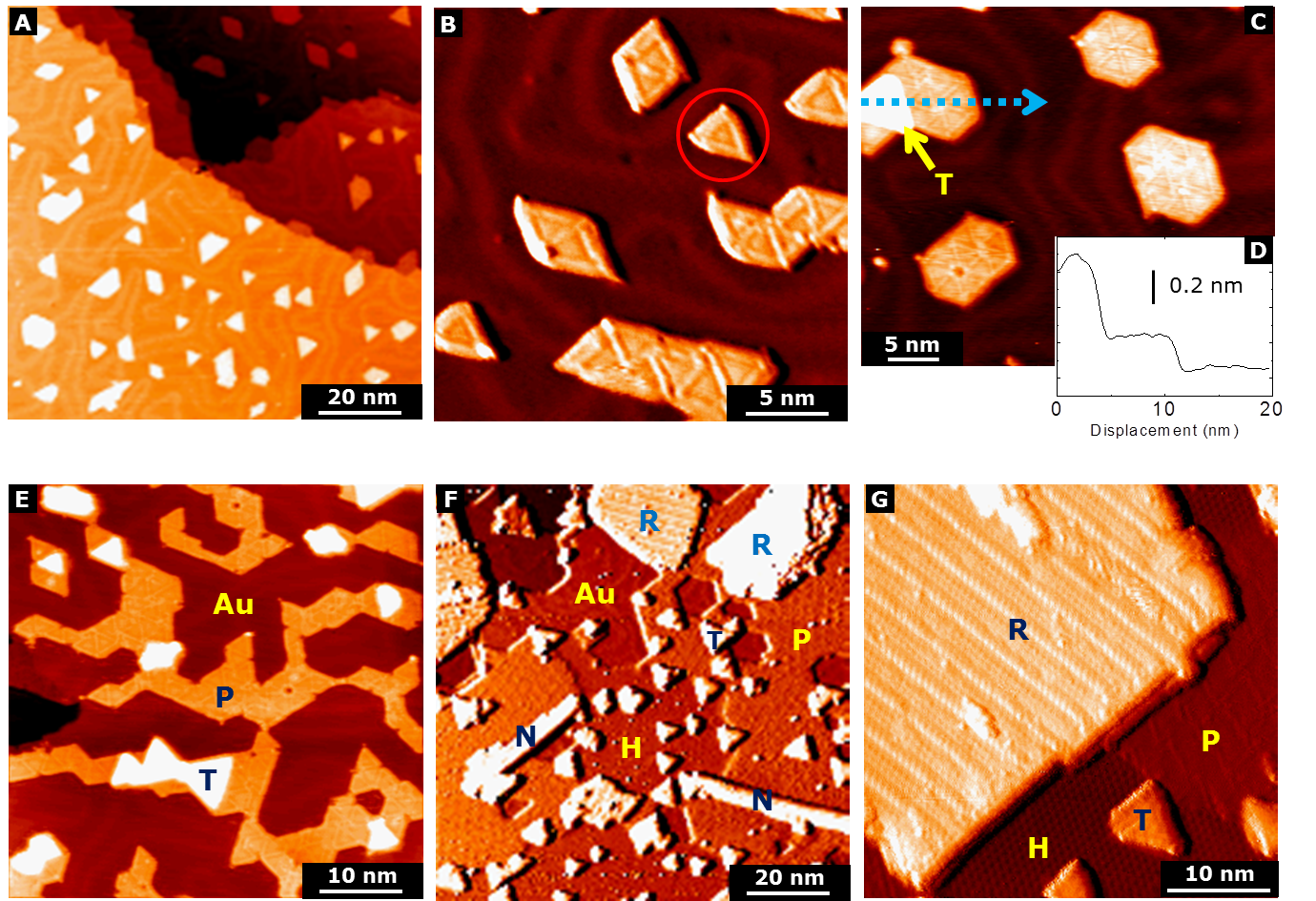}
\caption{STM images at different coverages. (b), (f) and (g) have been filtered to enhance the surface corrugation. (a) 0.25 ML ($\text{V}=-1.6 \,\text{V,}\:\; \text{I}=1 \,\text{nA}$); (b) 0.25 ML, a single triangular crystallite is circled ($\text{V}=-2.1 \,\text{V,}\:\; \text{I}=1 \,\text{nA}$); (c) 0.4 ML, yellow arrow points to an island of the triangular phase ($\text{V}=0.7 \,\text{V,}\:\; \text{I}=0.4 \,\text{nA}$); (d) topographic profile along the blue line in (c). (e) 0.65 ML ($\text{V}=0.9 \,\text{V,}\:\; \text{I}=0.5 \,\text{nA}$); (f) 1.5 ML ($\text{V}=1 \,\text{V,}\:\; \text{I}=0.4 \,\text{nA}$); (g) 1.5 ML ($\text{V}=0.6 \,\text{V,}\:\; \text{I}=0.7 \,\text{nA}$). Legend: Au = gold surface, P = pinwheel, T = triangular, H = honeycomb, R = row, N = needle.}
\label{fig.1}
\end{figure}
\\
\indent The evolution of the system in the investigated coverage range is shown in the STM images in fig. 1. At 0.25 ML (fig. 1a) we observe the formation of polygonal crystallites whose sides are aligned along directions forming $60^\circ$ or $120^\circ$ between each other. Some of these crystallites can also be observed to decorate the step edges of the substrate surface. The herringbone reconstruction is visibly perturbed by the presence of TiO\sub{x} crystallites that cause the distortion of the usual pattern of the ridges. Higher magnification images (fig. 1b) reveal that the polygonal crystallites are composed of smaller triangular crystallites (red circle in fig. 1b) merging together into larger islands. As the coverage increases (fig. 1c) TiO\sub{x} crystallites increase their size and most of them assume a peculiar hexagonal morphology, qualitatively similar to the pinwheel structures already reported for several ultrathin oxide films on metal substrates (e.g. VO\sub{x}/Pd(111) \cite{surnev}, TiO\sub{x}/Pt(111) \cite{sedona3}). Hence, in the following, we will refer to this kind of TiO\sub{x} structure as pinwheel phase. Pinwheel crystallites have occasionally been observed to show a variable STM contrast, depending on the bias voltage between tip and sample. However, although we did not thoroughly investigated such dependence, most of our measurements reveal an apparent height with respect to the substrate of $\sim 0.2$ nm (fig. 1d). In fig. 1c we also observe the formation of triangular islands on top of pinwheel crystallites (yellow label in fig. 1c): these islands, showing an apparent height of $\sim 0.65$ nm with respect to the substrate (fig. 1d), are morphologically different from the aforementioned triangular crystallites forming the pinwheel structure, and thus represent a different TiO\sub{x} phase that we refer to as \emph{triangular}. As the coverage increases above 0.5 ML (fig. 1e) pinwheel crystallites coalesce, starting to form a film that covers an increasingly larger portion of the Au surface, while triangular islands increase both in number and dimension. For coverage above 1 ML (fig. 1f-g), the Au surface is nearly completely covered by TiO\sub{x} structures. In this coverage range, we reveal the coexistence of five different TiO\sub{x} phases distinguished on the basis of their morphology. The pinwheel and honeycomb phases appear as well-ordered monolayer films wetting uniformly the Au surface. In addition to triangular islands, two other kinds of structures, namely \emph{row} and \emph{needle}, are revealed in STM images.

\subsection{Structural properties of the honeycomb phase}
With the aim of properly interpreting the structure of the honeycomb phase, the experimental results provided by STM and LEED measurements were compared with those predicted by a theoretical investigation based on dedicated DFT calculations of the electronic structure. This approach has been proved to be valuable for the identification of the structural model in best agreement with the experimental data of similar honeycomb oxide structures (e.g. TiO\sub{x} \emph{kagomé} structure on Pt(111) \cite{barcaro}, VO\sub{x} honeycomb on Pd(111) \cite{surnev}).
\begin{figure}[htbp]
\centering%
\includegraphics[scale=0.6]{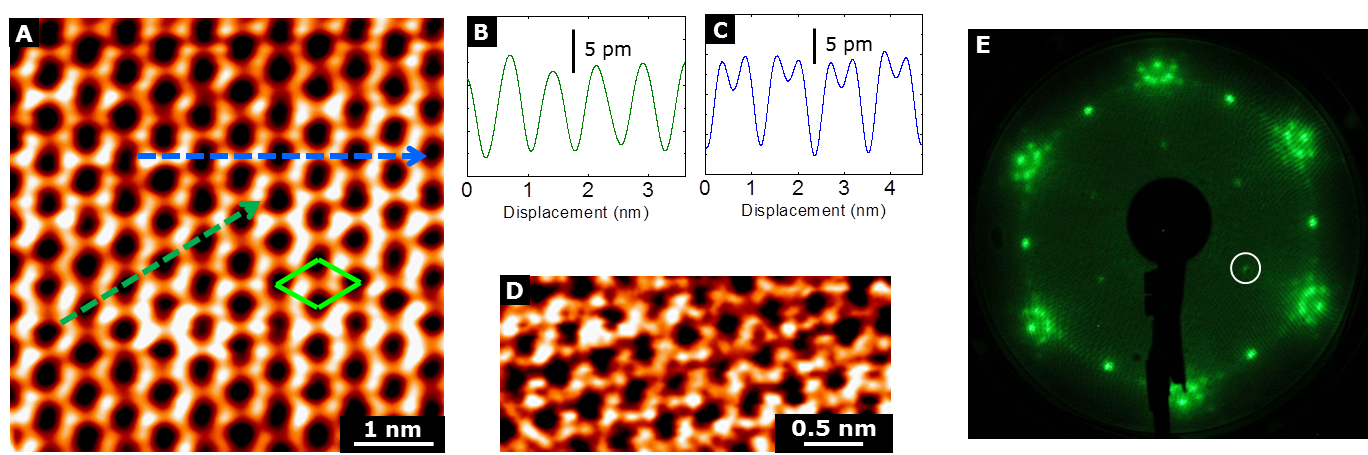}
\caption{(a) STM image of the honeycomb phase, 1.5 ML ($\text{V}=0.6 \,\text{V,}\:\; \text{I}=0.2 \,\text{nA}$). (b) and (c) respectively report topographic profiles along the green and blue lines in (a). (d) STM image showing the kagomé structure, 1.5 ML ($\text{V}=0.3 \,\text{V,}\:\; \text{I}=0.7 \,\text{nA}$). (e) LEED pattern at 1.5 ML (50 eV). White circle indicates a $(2\times 2)$ spot.}
\label{fig.2}
\end{figure}
\\
\indent The honeycomb layer exhibits an apparent height of $\sim 1$ {\AA} with respect to the underlying Au surface. In fig. 2a, a high-resolution STM image shows the peculiar honeycomb geometry characterizing this phase. The unit cell is displayed on the image and two topographic profiles are reported: the green one (fig. 2b) reveals a lattice periodicity of $\sim 6$ {\AA}, the other one (blue line, fig. 2c) reveals a finer structure characterized by a $\sim 4$ {\AA} interatomic distance. Occasionally, probably under particular conditions of the tip apex state, a \emph{kagomé} structure was observed instead of the usual honeycomb arrangement (fig. 2d). The LEED pattern in fig. 2e reveals the presence of a $(2\times 2)$ superstructure aligned along the symmetry directions of Au(111)-$(1\times 1)$, in agreement with STM measurements. As will be clarified in the following section, the satellite spots near the $(1\times 1)$ spots can be attributed to the reciprocal space contributions of the pinwheel superstructure. 
\begin{figure}[htbp]
\centering%
\includegraphics[scale=0.6]{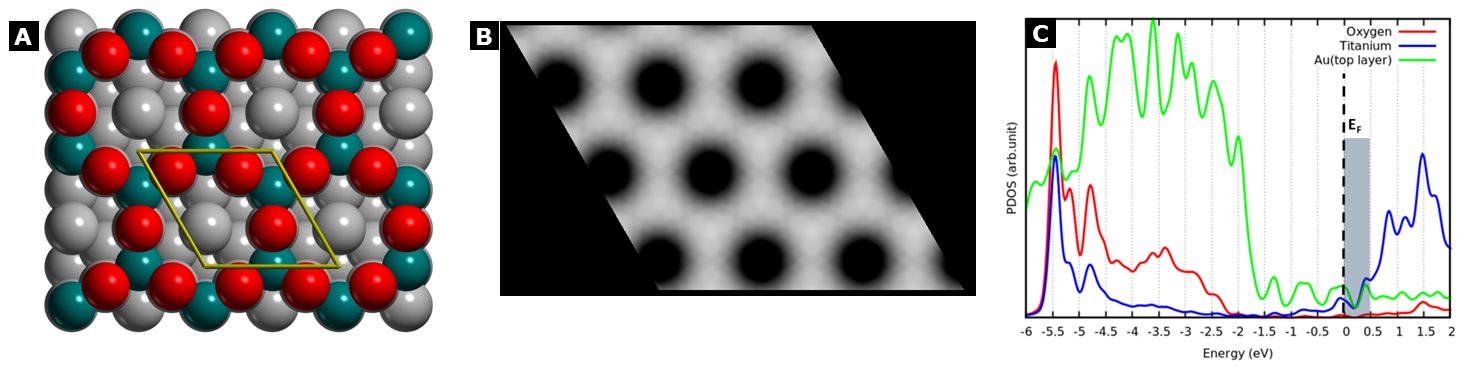}
\caption{(a) Top view of the honeycomb Ti\sub{2}O\sub{3}/Au structure centered on top Au atoms (grey = Au, green = Ti, red = O). (b) Simulated STM image at 0.5 V. (c) Projected density of states showing he different contributions from Au, Ti, and O atoms. The shadowed energy window between the Fermi level and 0.5 eV is used to obtain the simulated STM image in (b).}
\label{fig.3}
\end{figure}
\\
\indent To interpret the honeycomb structure, Wu et al. \cite{wu} suggested a structural model based on a Ti-O bilayer with Ti atoms directly in contact with the gold surface, forming a $(2\times 2)$ superstructure with Ti\sub{2}O\sub{3} stoichiometry. In particular, in their model Ti atoms occupy hollow sites of Au(111)-$(1\times 1)$ to give a honeycomb arrangement centered on top Au atoms, while oxygen atoms occupy bridge position between two adjacent Ti atoms. Our theoretical study was conducted by initially considering such model (fig. 3a) and another one, based on the same Ti-O lattice but centered on HCP sites of the gold surface. The former is found to be more stable, with a relevant energy gain of 0.14 eV. The simulated STM image, shown in fig. 3b, was generated integrating the calculated local density of states in the energy interval between the Fermi level (0 eV) and 0.5 eV. In this range, the image contrast is mainly due to Ti atoms, as can be seen from the partial density of states plotted in fig. 3c, while the contribution to the tunneling current of oxygen electronic states is negligible in the typical energy range of STM measurements. The good agreement between simulated and experimental STM data provides evidence for the accuracy of the proposed model. Incidentally, this structural model can also account for the observation of the kagomé arrangement as due to an occasional imaging of oxygen atoms due to an unusual tip state. 

\subsection{Structural properties of the pinwheel phase}
Among the TiO\sub{x} phases that we identified, the pinwheel is the only one that is always present in the investigated coverage range, under our experimental conditions. However, it is important to point out that the morphology of this phase evolves significantly from the low to the high coverage regime. At low coverage, i.e. below 0.5 ML, we observed single triangular crystallites (see fig. 1b) starting to merge together into more extended crystallites which exhibit the surface motif of the pinwheel phase (fig. 4a). At higher coverage, i.e 0.5 - 1 ML, we observed the coalescence of pinwheel crystallites into a continuous film which gradually wets the substrate surface (fig. 4b). Hence, our measurements reveal the formation of the pinwheel layer starting from single triangular crystallites which act as elementary ``building units''. The high-resolution STM image in fig. 4c reveals the typical surface appearance of the pinwheel film: the hexagonal arrangement of the ``hubs'' gives rise to a superstructure with a periodicity of $\sim 30$ {\AA}. From the analysis of our LEED measurements, the pinwheel superstructure can be described as a $(4\sqrt{7}\times 4\sqrt{7})R19.1^\circ$  commensurate coincidence lattice. The reciprocal lattice (white dots in fig. 5d), calculated from the direct lattice in fig. 4e, is in agreement with the experimental LEED pattern. The double satellite spots arise from two differently oriented domains.  
\begin{figure}[htbp]
\centering%
\includegraphics[scale=0.6]{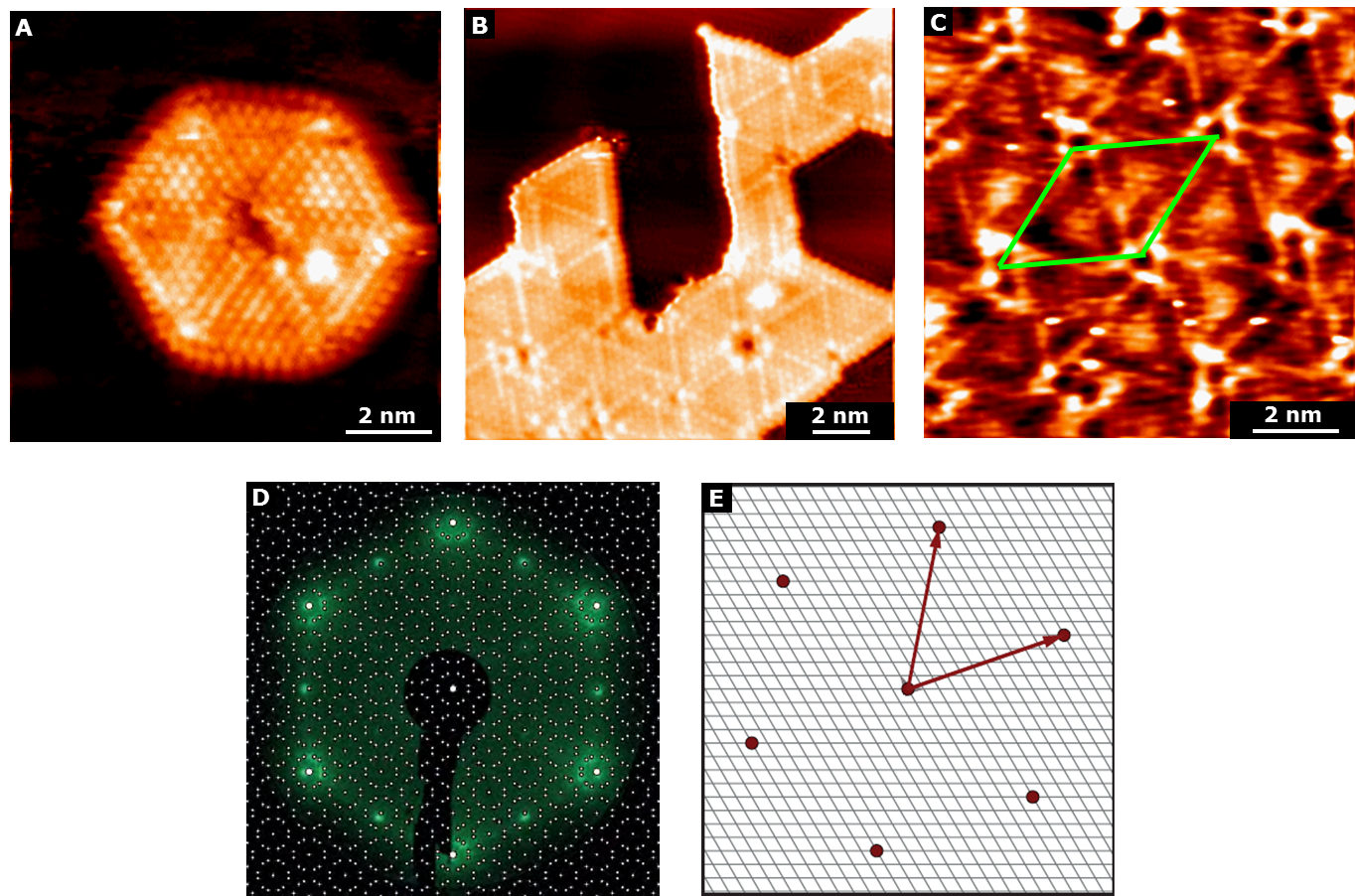}
\caption{(a) STM image at 0.3 ML showing a hexagonal pinwheel crystallite ($\text{V}=0.1 \,\text{V,}\:\; \text{I}=2.5 \,\text{nA}$). (b) STM image at 0.7 ML showing the formation of a continuous pinwheel film ($\text{V}=1.5 \,\text{V,}\:\; \text{I}=0.4 \,\text{nA}$). (c) STM image of the pinwheel surface at 1.5 ML ($\text{V}=0.1 \,\text{V,}\:\; \text{I}=1.5 \,\text{nA}$). (d) LEED pattern at 1.5 ML (50 eV). White dots indicate reciprocal space points obtained from the direct lattice in (e). (e) Direct space representation (grey = Au(111)-$(1\times 1)$ lattice, red = pinwheel superstructure).}
\label{fig.4}
\end{figure}
\\
\indent A possible interpretation of the pinwheel structure can be related to the formation of a moiré coincidence pattern due to the mismatch between the substrate and the overlayer. In fig. 5 an hexagonal lattice of Ti atoms (red dots) is superimposed to the substrate $(1\times 1)$ lattice (blue dots). In order to reproduce the $(4\sqrt{7}\times 4\sqrt{7})R19.1^\circ$  superstructure, the Ti lattice is rotated by $2.11^\circ$ with respect to the substrate and has a 3.43 {\AA} interatomic distance, which agrees with the value measured from atomic resolution STM images, i.e. $3.2\pm 0.2$ \AA. Following the common assumption that the STM contrast of a titanium oxide surface at positive bias is dominated by the tunneling contributions of Ti atoms \cite{diebold}, we can tentatively approach the interpretation of the pinwheel contrast on the basis of a purely topographic modulation of the Ti layer due to the atoms occupancy of different substrate sites, i.e. hollow, bridge and on-top sites. In the limit of this hypothesis, the atomistic representation in fig. 5 can be considered effective in qualitatively reproducing the enhanced contrast of the ``spokes'' and the triangular areas between them: bright areas largely correspond to bridge or on-top Ti sites, while darker ones to hollow or quasi-hollow Ti sites. Nonetheless it should be noticed that this interpretation fails when considering the coincidence sites, where the local electronic structure can be expected to play a leading role in the resulting contrast. A proper interpretation of the STM contrast clearly requires accurate knowledge of the surface electronic structure, which is of fundamental importance in determining the spatial variations of tunneling current. However, the proposed coincidence model can be reasonably considered as a starting point to build a more advanced structural model, capable to account for possible interaction effects with the substrate and for the influence of oxygen atoms on the spatial distribution of electronic states.\\
\indent A qualitatively similar interpretation of the pinwheel structure based on the moiré model was suggested also in the work of Wu et al. \cite{wu}. However, our experimental data based on STM and LEED measurements are not in agreement with those derived from the model they proposed, which produces a $(\sqrt{67}\times\sqrt{67})R12.2^\circ$ superstructure with a 23 {\AA} periodicity.
\begin{figure}[htbp]
\centering%
\includegraphics[scale=0.4]{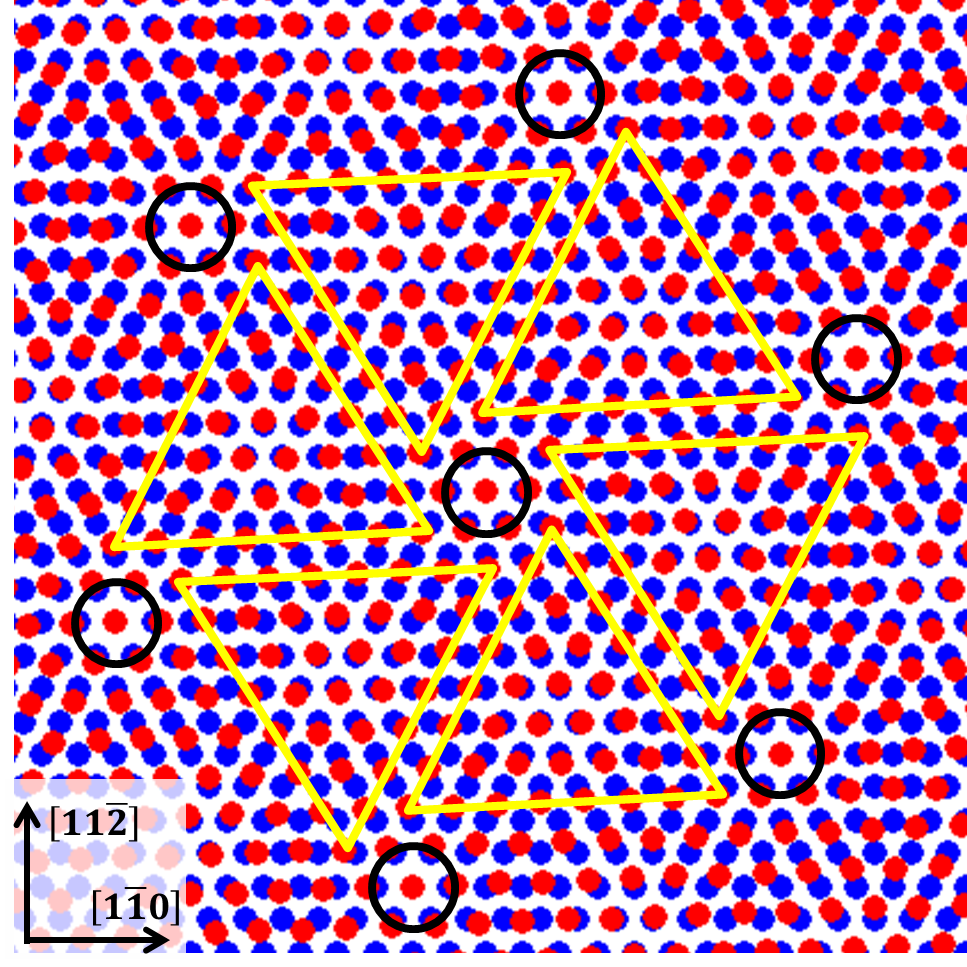}
\caption{Atomistic model of the moiré coincidence pattern. Ti atoms are arranged in hexagonal lattice (red dots, 3.43 {\AA} periodicity) rotated by $2.11^\circ$ with respect to the Au lattice (blue dots, 2.89 {\AA} periodicity). The resulting coincidence site (black circles) represent a $(4\sqrt{7}\times 4\sqrt{7})R19.1^\circ$  superstructure. Yellow lines highlight the triangular areas between the ``spokes'' of the pinwheel.}
\label{fig.5}
\end{figure}

\subsection{Structural properties of the triangular, row, and needle phases}
As anticipated when discussing the evolution of the TiO\sub{x}/Au(111) system, the triangular phase has been observed to grow on top of pinwheel crystallites starting from coverages around 0.4 ML. As the coverage increases, triangular islands are observed on top of both pinwheel and honeycomb layers. The average side length is about 5 nm and most of the islands have an apparent height of $0.65\pm 0.03$ nm with respect to the bare Au surface (e.g. see the topographic profiles in fig. 6c or 1d). Occasionally, especially at high coverages (i.e. above 1 ML), we observed larger islands having an apparent height of $\sim 0.9$ nm with respect to the Au surface (fig. 6a-c). Since no triangular islands were observed with an apparent height much lower than 0.65 nm and none of these measurements showed a significant dependence on the applied bias, our observations suggest triangular islands to be formed by a stable $\sim 0.65$ nm (with respect to the Au surface) base structure that grows by adding subsequent layers $0.25\pm 0.02$ nm thick. This hypothesis seems to be supported by high-magnification STM images (fig. 6d) showing the stacking of different layers. The results of our structural characterization of the triangular phase appear to be compatible with those reported by Potapenko et al. \cite{pota} which suggest those structures could be based on rutile-TiO2(100) layers, whose thickness is known to be 0.23 nm.
\begin{figure}[htbp]
\centering%
\includegraphics[scale=0.6]{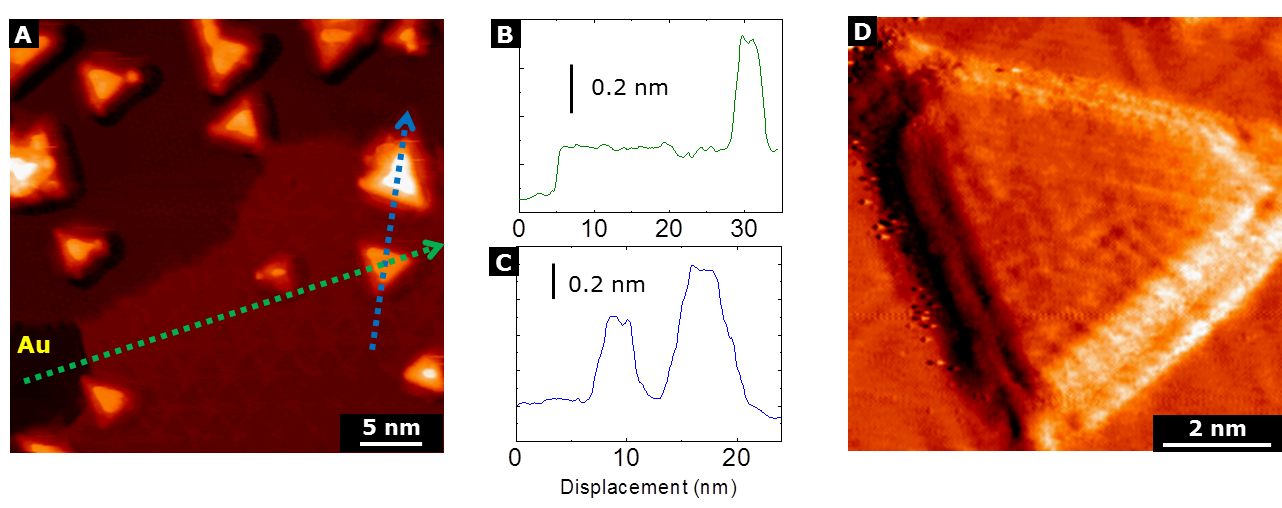}
\caption{(a) STM image showing triangular islands on top of pinwheel and honeycomb films, 1.5 ML ($\text{V}=0.8 \,\text{V,}\:\; \text{I}=0.4 \,\text{nA}$). (b) Topographic profile along the green line in (a) showing the apparent height of a triangular island with respect to the Au surface, i.e. $\sim 0.65$ nm. (c) Topographic profile along the blue line in (a) showing the height difference between two triangular islands, i.e. $\sim 0.25$ nm. (d) Filtered STM image showing the layered structure of a triangular island.}
\label{fig.6}
\end{figure}
\\ 
\indent An island of the \emph{row} phase is reported in the STM image in fig. 7a, showing the surface morphology. The linear size of these islands usually ranges from 20 nm to 50 nm. The characteristic rows are oriented along the $[11\overline{2}]$ direction of the substrate and different orientation domain can be observed on the same island forming angles of $120^\circ$ between each other. The island shape is usually irregular, although occasionally a preferential orientation of their sides along the $[11\overline{0}]$ direction can be observed, as shown in fig. 7a. Similarly to the triangular phase, height measurements suggest the row phase to be characterized by a base layer $\sim 0.65$ nm high (with respect to the Au surface) and subsequent layers $\sim 0.25$ nm high (fig. 7b-c). From atomic resolution images (fig. 7d) it is possible to observe the regular alternation of brighter and darker atomic rows. The size of the rectangular unit cell is $5.7\pm 0.2$ {\AA} $\times\: 29\pm 0.5$ {\AA}. The morphology of the so-called \emph{star} phase observed by Ragazzon et al. \cite{raga1} seems to be approximately compatible with our STM measurements of the row phase. The \emph{star} phase was hypothesized to be related to a $8\times 1$ reconstruction of rutile-TiO\sub{2}(100).\\
\begin{figure}[htbp]
\centering%
\includegraphics[scale=0.6]{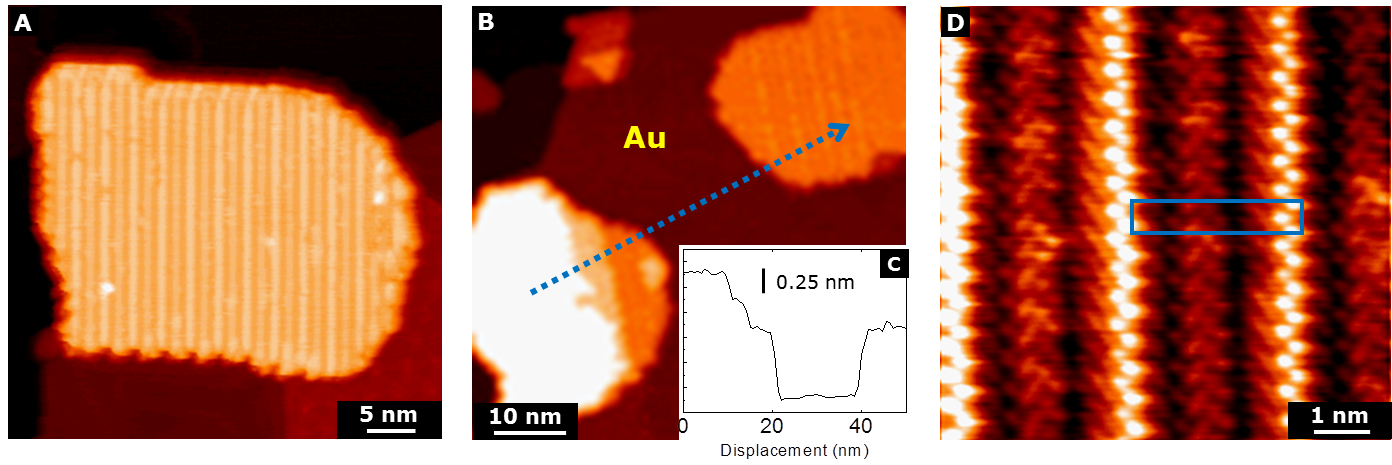}
\caption{(a) STM image of an island of the row phase (1 ML, $\text{V}=0.8 \,\text{V,}\:\; \text{I}=0.2 \,\text{nA}$). (b) STM image showing different layers on the same island (1.5 ML, $\text{V}=1.5 \,\text{V,}\:\; \text{I}=1 \,\text{nA}$). (c) Topographic profile along the blue line in (b). (d) Atomically resolved STM image of the row structure (1 ML, $\text{V}=0.3 \,\text{V,}\:\; \text{I}=2.3 \,\text{nA}$). The blue box represents the rectangular unit cell.}
\label{fig.7}
\end{figure}
\indent Our STM observations also allowed to reveal needle-like crystallites up to 150 nm long, having typically a width of few nanometers ($\sim 5$ nm). In fig. 8a we report a high-resolution STM image showing part of a needle crystallite and a cross-section profile (fig. 8b) highlighting the non-planar morphology of this structure, which qualitatively resembles the shape of a ``nanowire''. From atomic resolution images (fig. 8c) it is possible to measure a $3.2\pm 0.2$ {\AA} interatomic distance along the main direction (fig. 8d) and a $5.2\pm 0.1$ {\AA} spacing between adjacent atomic rows. Incidentally, from previous STM measurements on rutile-TiO\sub{2}(110) we found a $3.2\pm 0.2$ {\AA} interatomic distance along [001] oriented atomic rows, compatible with the atomic periodicity of the needle phase.
\begin{figure}[htbp]
\centering%
\includegraphics[scale=0.6]{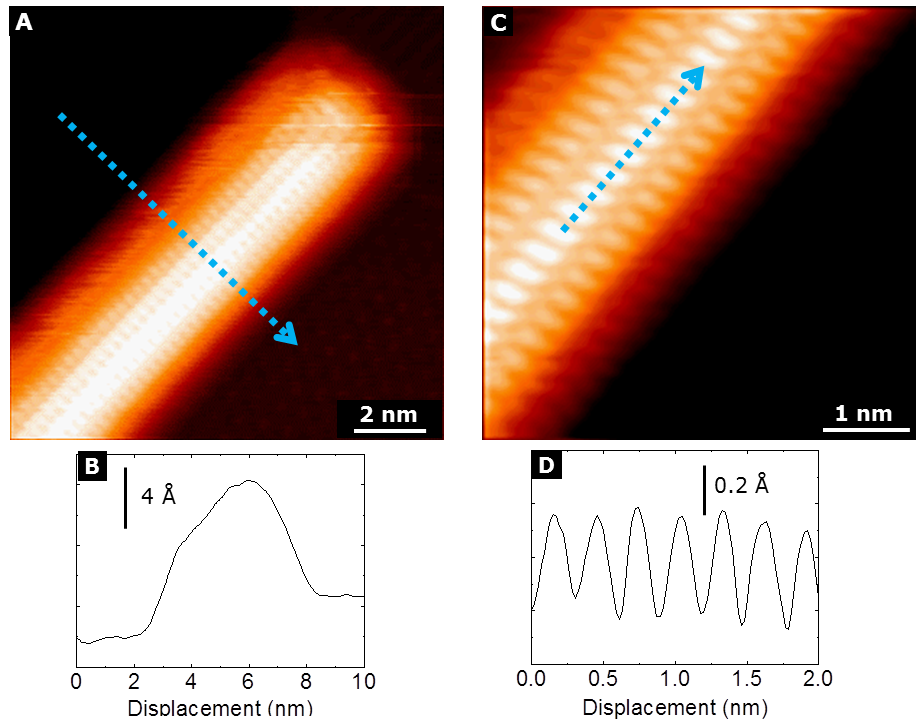}
\caption{(a)-(b) Atomically resolved STM images of the needle structure (1 ML, $\text{V}=0.1 \,\text{V,}\:\; \text{I}=3.1 \,\text{nA}$). (c) Topographic profiles along the line in (a). (c) Topographic profiles along the line in (b) showing the atomic periodicity.}
\label{fig.8}
\end{figure}

\subsection{Electronic properties of the TiO\sub{x} phases}
In order to explore the electronic properties of the different phases, we performed STS measurements of the observed structures. This investigation does not include the STS characterization of needle crystallites, since they were observed less frequently in comparison to the other four phases. \\
\indent Figure 9a shows the normalized dI/dV curve obtained for the honeycomb phase in an energy interval centred on the Fermi level. The main characteristics of the STS spectrum are a pronounced DOS in the empty states region (i.e. above the Fermi energy), a minimum around the Fermi level and smaller DOS contributions in the occupied states region. Interestingly, the differential conductivity, hence the electronic DOS, does not drop to zero around the Fermi level, probably suggesting a partial metallization of the oxide layer. The theoretical investigation of the honeycomb structure provides useful support in assessing the nature of the experimentally observed electronic features. The calculated DOS of the honeycomb/Au(111) system is shown in fig. 9b. 
The features at positive energy (i.e. virtual states) in the range 1-2 eV above the Fermi level are mainly due to Ti-3d orbitals. At negative energies (occupied orbitals), between -0.5 and -2 eV with respect to the Fermi level, Au(111) surface states are present, while contributions due to O and Ti ions almost vanish in this region. This suggests to interpret the feature at positive energy in the STS spectrum as a Ti-state, while the peak at negative energy should be attributed to gold surface states. A quantitative comparison between density of states obtained from DFT calculations and STS spectra is, however, difficult, due to three aspects. First of all, the position of the Ti\sub{2}O\sub{3} orbitals with respect to the Fermi level is affected by an intrinsic limit of the GGA approximation, i.e. the so-called self-interaction error \cite{dago}, which reflects in an underestimation of the electron band gap of materials such as TiO\sub{2}. Moreover, the intensity of the STS peaks depends on several factors controlling the tunnelling probability, while a DOS plot reports all eigenstates of the system. Also the sample-tip distance in the measurements may introduce discrepancies with the calculated DOS. We tried to take this latter factor into account by calculating also the DOS projected onto a uniform layer of chargeless spheres located 2 {\AA} above the apical ions of the surface. As shown in fig. 9c, the density of states above the surface reflects all the main features of the DOS projected on the ions, but the contribution of the Au orbitals is smaller. Finally, when dealing with Au(111), a peculiar feature such as the herringbone reconstruction should be taken into account. As shown in previous reports, the position of the gold surface states depends remarkably on this reconstruction \cite{lauwaet}.\\
\begin{figure}[htbp]
\centering%
\includegraphics[scale=0.4]{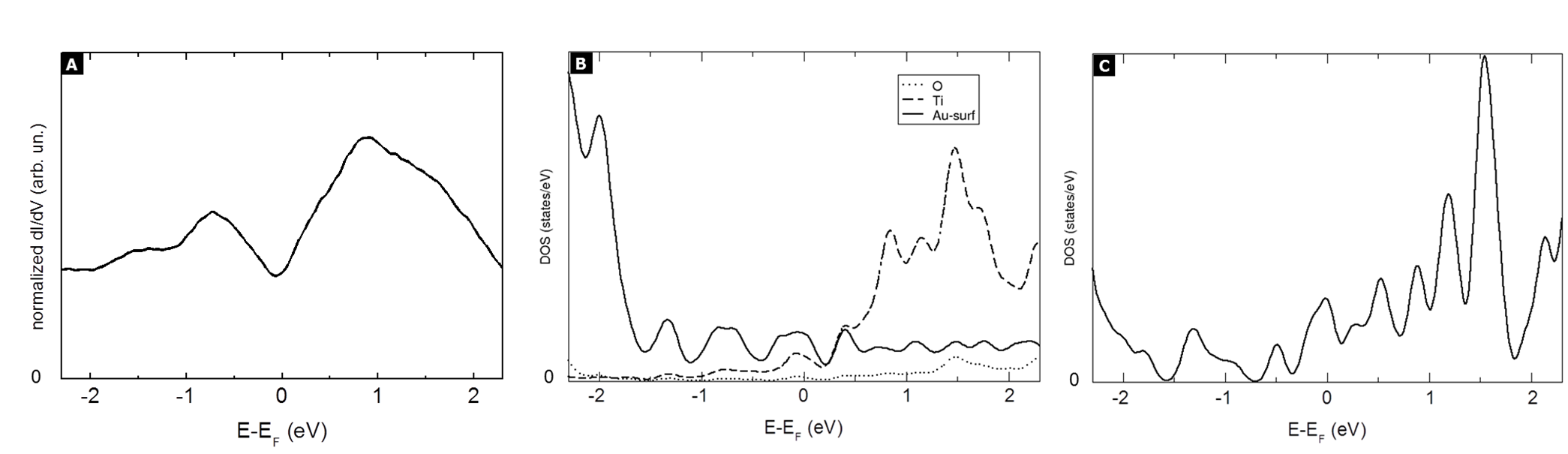}
\caption{Electronic properties of the honeycomb phase. (a) Normalized experimental dI/dV curve. (b) Calculated density of states of the honeycomb/Au(111) system. (c) Density of states calculated at 2 Å from the surface.}
\label{fig.9}
\end{figure}

\begin{figure}[htbp]
\centering%
\includegraphics[scale=0.42]{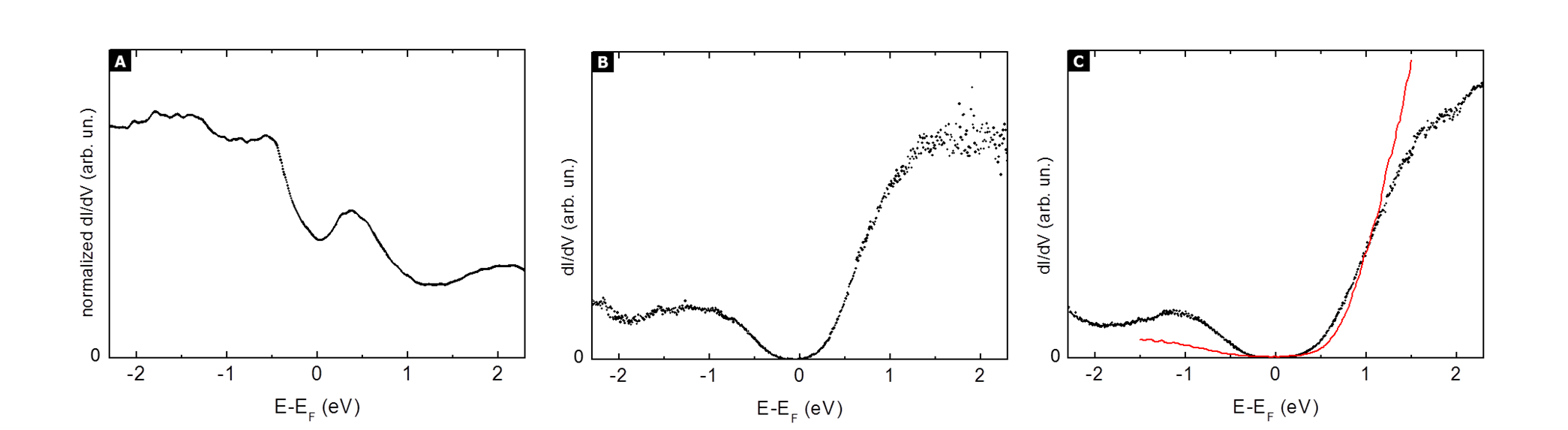}
\caption{Representative STS curves for different TiOx phases. (a) Normalized experimental dI/dV curve of the honeycomb phase (black dots) and calculated DOS (solid red line). (b) Normalized dI/dV curve of the pinwheel phase. (c) dI/dV curve of the triangular phase. (d) dI/dV curve of the row phase (black dots) and of rutile-TiO\sub{2}(110) (solid red line).}
\label{fig.10}
\end{figure}

\indent Figure 10a shows the normalized dI/dV curve of the pinwheel phase. Around the Fermi level the behaviour is similar to the honeycomb spectrum, suggesting that a partial metallization of the oxide film occurs also for the pinwheel phase. Our analysis reveals a filled states band right below the Fermi level and two DOS features, respectively at $\sim 0.4$ and $\sim 2.1$ eV in the empty states. From a qualitative perspective, it may be interesting to notice that, in contrast to the honeycomb DOS behaviour (fig. 9a), the normalized dI/dV of the pinwheel phase is characterized by a lower density in the empty states region with respect to the occupied states region. A tentative interpretation of this behaviour could be based on the hypothesis that the pinwheel structure is characterized by a lower Ti oxidation state which results in a lower density of empty states (under the assumption that Ti 3d states largely contribute to the density of unoccupied states). In this respect, it may be worth mentioning that in the case of TiO\sub{x} pinwheel observed on Pt(111) DFT calculations supported a structural model characterized by an overall TiO\sub{1.2} stoichiometry with a Ti\up{+2.4} formal oxidation state \cite{barcaro}, that is lower than the Ti\up{+3} oxidation state derived from the honeycomb model.\\ 
\indent Figures 10b-c show the differential conductivity curves acquired for the triangular (fig. 10b) and row (fig. 10c, black dots) phases. In this case the application of a normalization procedure to the experimental curves was not considered to be important since the typical exponential behaviour of the tunneling transmission coefficient does not hinder the identification of the main electronic characteristics. The two experimental curves are remarkably similar, showing both a broad feature around $-1$ eV, zero DOS at the Fermi level, and a considerable conductivity increase in the positive bias region. The overall behaviour is significantly different from monolayer TiO\sub{x} phases, i.e. pinwheel and honeycomb, and can be qualitatively compared to typical STS spectra of reduced TiO\sub{2} surfaces, such as the one measured on rutile-TiO\sub{2}(110) shown in fig. 10c (red line). This aspect is in agreement with the aforementioned hypotheses (see Sec. 3.3) that the structure of these phases can be related to a termination of a bulk TiO\sub{2} phase.

%% file: conclusions.tex
\section{Conclusions}
We studied the evolution of titanium oxide 2D nanostructures supported by the Au(111) surface, for increasing coverage up to 1.5 ML. Titanium oxide has been obtained after Ti evaporation onto the substrate, followed by O\sub{2} exposure and annealing. We characterized the morphological, structural and electronic properties of the obtained structures by means of STM, STS and LEED measurements. Our experiments allowed to identify five morphologically distinct TiO\sub{x} phases. The honeycomb phase appears as a monolayer film giving rise to a $(2\times 2)$ superstructure. On the basis of the results of DFT calculations, a structural model characterized by a Ti\sub{2}O\sub{3} stoichiometry has been discussed and supported. The pinwheel phase was observed both at low ($< 0.5$ ML) and high ($> 1$ ML) coverage, evolving from single crystallites to a continuous ordered monolayer film characterized by a $(4\sqrt{7}\times 4\sqrt{7})R19.1^\circ$ superstructure. A moiré model has been shown to accurately reproduce the superstructure cell and the interatomic distance, and thus can be considered as a starting point for a more advanced modeling of the pinwheel structure. The triangular and row phases both showed a layered structure with a layer height of $\sim 0.25$ nm. Atomic resolution STM images of the row phase allowed to measure a $5.7\pm 0.2$ {\AA} $\times \: 29\pm 0.5$ {\AA} rectangular unit cell. In addition, we reported atomic resolution images of the needle phase which appears in the form of needle-like crystallites showing a ``nanowire'' morphology, whose structural properties are still unexplored. STS measurements allowed to acquire information on the electronic properties of the observed TiO\sub{x} structures. As far as honeycomb and pinwheel phases are concerned, the results of our analysis suggest a partial metallization of the oxide layer, probably resulting from a significant interaction with the substrate. The differential conductivity curves measured for the triangular and row phases show a similarity with those obtained on bulk TiO\sub{2} surfaces, thus encouraging the hypothesis that the structure of these phases can be related to a termination of a bulk TiO\sub{2} phase. The experimental and theoretical results that we presented provide insight into the structural and electronic properties of titanium oxide 2D nanostructures supported on a gold substrate. In particular, monolayer structures such as honeycomb and pinwheel show metallic-like character with a non-zero DOS at the Fermi level, while semiconducting behaviour is shown by triangular and row. The observed change of electronic properties resulting from different 2D structural organizations that are not allowed in bulk systems makes titanium oxide another important metal oxide candidate in the prominent family of 2D oxide materials.\\

\paragraph{Acknowledgements.}
We are grateful to Prof. Gianfranco Pacchioni for his valuable collaboration, useful discussion and critical reading of this manuscript.